\documentclass[aps,prd,twocolumn,nofootinbib]{revtex4-1}
\usepackage[T1]{fontenc}
\usepackage[utf8]{inputenc}
\usepackage[english]{babel}
\usepackage{url}
\usepackage{mathrsfs}
\usepackage{amsfonts}
\usepackage{amsmath}
\usepackage{amssymb}
\usepackage{amsthm}
\usepackage{graphicx}

\usepackage[toc,title]{appendix}
\usepackage{bm}
\usepackage{booktabs}
\usepackage{braket}
\usepackage{caption}
\usepackage{enumerate}
\usepackage{enumitem}
\usepackage[Bjornstrup]{fncychap} 
\usepackage{fnpos} %
\usepackage[bottom,multiple]{footmisc} 
\usepackage{lipsum}	
\usepackage{mathtools} 
\usepackage[]{natbib}	
\usepackage{quoting}
\quotingsetup{font=small}
\usepackage{sidecap}
\usepackage{subfig}
\usepackage{textcomp}
\usepackage{tikz}
\usetikzlibrary{decorations.pathmorphing,calc}
\usepackage{titlesec}
\usepackage{tensor}
\usepackage{xfrac}

\usepackage{csquotes}
\usepackage{helvet}
\usepackage{listings}

\usepackage[bookmarks,bookmarksopen=true,colorlinks,citecolor=blue,linkcolor=blue,urlcolor=blue]{hyperref}

\makeFNbelow
\newcommand{\be}{\begin{equation}}
\newcommand{\ee}{\end{equation}}
\newcommand{\bn}{\begin{equation*}}
\newcommand{\en}{\end{equation*}}

\newcommand{\lie}{\pounds} 
\newcommand{\Ae}{\text{\AE}} 
\newcommand{\ux}{(u\cdot \xi)}

\newcommand{\ax}{(a\cdot \xi)}
\newcommand{\ut}{(u\cdot t)}
\newcommand{\st}{(s\cdot t)}
\newcommand{\as}{(a\cdot s)}

\newcommand{\M}{\mathcal{M}}

\newcommand{\ru}{r_{\text{UH}}}

\newcommand{\tu}{T_\text{UH}}

\newcommand{\ku}{\kappa_{\text{UH}}}
\newcommand{\kk}{\kappa_{\text{KH}}}
\newcommand{\su}{S_\text{UH}}
\newcommand{\ou}{\Omega_\text{UH}}
\newcommand{\K}{\tensor[^2]{K}{}}
\newcommand{\Ha}{\mathbb{H}}
\newcommand{\Ma}{\mathbb{M}}
\newcommand{\Ja}{\mathbb{J}}
\newcommand{\B}{\mathcal{B}_T}

\theoremstyle{plain}

\begin{document}

\rmfamily

\title{On the first law of black holes with a universal horizon}
\date{\today}
\author{Costantino Pacilio}
\author{Stefano Liberati}
\affiliation{SISSA, Via Bonomea 265, 34136 Trieste, Italy, EU}
\affiliation{INFN, Sezione di Trieste, Trieste, Italy}
\begin{abstract}
In Lorentz violating theories of gravitation with a preferred foliation a notion of black hole is still possible, despite the presence of infinitely fast propagating modes. Such event horizons are known as universal horizons. Their discovery poses the question of whether they satisfy mechanical laws, analogous to the ones of Killing horizons in Lorentz symmetric theories, and whether they admit a thermodynamical interpretation. In this paper we study the viability of the first law for several exact universal horizon solutions previously derived in the literature. Our results show that a simple mechanical and thermodynamical interpretation is problematic in these cases, and call for a more systematic study of rotating universal horizons.
\end{abstract}
\maketitle
%
\section{Introduction}
In general relativity (GR) a black hole (BH) is defined as a region causally disconnected from spatial infinity. The notion of causality is provided by the light cones, and is rooted into the property that physical modes cannot travel faster than light, i.e. in local Lorentz symmetry (LLS).

If we consider a gravity theory in which LLS is broken, we can ask what the fate is of the BH concept. In the case that the species have different, but finite, limiting speeds, one can still define different BHs for each species. Instead, when the modes have unlimited speed, it may seem that it is impossible to define a BH. This conclusion is incorrect.

In fact, in theories where LLS is broken by the introduction of a preferred foliation, there is a proper notion of BH. In these theories causality is defined by the requirement that causal modes move forward with respect to the preferred time direction. In general the preferred  leaves extend up to spatial infinity; but when a leaf does not satisfy this property, it bounds a region disconnected from spatial infinity. In other words, BHs are the effect of the relative deformation of the leaves. For a formal treatment see \cite{uh_global}.

These Lorentz violating (LV) BHs were first discovered \cite{blas_uh, barausse_black} in the context of static solutions of (the infrared version of) Ho{\v r}ava-Lifshitz (HL) gravity \cite{horava, horava_status, wang_review}, a modified theory of gravitation with a preferred foliation, and of Einstein-aether (\Ae) theory \cite{aether, jacobson:report}, in which LLS is broken by a preferred timelike vector field. The horizon determined by such BHs is dubbed universal horizon (UH). It is important to notice that the universal horizons generally exist alongside a Killing horizon (KH).

The theoretical discovery of universal horizons poses questions about their analogy with the more familiar Killing horizons. In particular, given that Killing horizons obey mechanical laws, and that these laws admit a thermodynamical interpretation \cite{wald_book}, it is natural to ask if similar laws hold also for universal horizons.

An important progress in this direction has been a tentative identification of a notion of "universal horizon temperature" $\tu$ \cite{mattingly_tunneling, cropp_ray, ding_radiation}. However, it is not clear to which extent $\tu$ is a temperature \cite{parentani}, or if mechanical and thermodynamical laws are associated to it: for example, a proof of the zeroth law has been given in \cite{uh_global} in a widely general setting, while the validity of a first law is still controversial.

The aim of the present paper is precisely to understand if black holes with UHs admit a first law of mechanics with a thermodynamical interpretation. This study was initiated in \cite{foster_noether}, before the discovery of universal horizons, using Wald's covariant Hamiltonian construction at the Killing horizon, and no physical interpretation of the first law emerged. After the discovery of universal horizons, it was suggested in \cite{mattingly:uh} that the first law ought to be associated with them.

Following this suggestion, we analyze explicitly a range of exact UH solutions, to see if they satisfy a first law in the form
\be
\label{eq:law:1}
d\text{(Mass)}=\tu\,d\su+\text{(work terms)}.
\ee
We interpret \eqref{eq:law:1} as a differential equation to be solved for $\su$. In particular this means that we do not assume \emph{a priori} that $\su$ is proportional to the area of the UH.

Clearly there is some vagueness, because we could always ascribe any extra term in \eqref{eq:law:1} to a not better specified form of work. For this reason we restrict ourselves to the simplest and most natural choices of work terms. In particular, since the static solutions we consider have a one parameter dependence, we assume in analogy with GR that no work term is involved. Similarly, when considering rotating solutions, we allow only for the work term due to the change in angular momentum. Moreover, we do not consider variations of the cosmological constant.

We will work with the infrared version of Ho{\v r}ava gravity, in which most of the known UH exact solutions have been obtained so far. More specifically, we focus on the healthy extension \cite{horava_healthy} in its covariant formulation \cite{jacobson:horava}, also known as khronometric theory or T-theory.

The paper is organized as follows. In Sec.\,\ref{sec:action} we review the action and the equations of motion of T-theory. In Sec.\,\ref{sec:hamiltonian}, following \cite{jacobson_donnelly, mattingly:lifshitz}, we derive the Hamiltonian of T-theory, from which we extract the definition of mass to be used in \eqref{eq:law:1}; in particular in Sec.\,\ref{sub:regularization} we introduce a regularization scheme for the Hamiltonian, which allows us to deal with the general case of nonvanishing asymptotic shift. In Sec.\,\ref{sec:law} we study the first laws of several exact UH solutions of T-theory, along the lines explained above: we find that, as the complexity of the solutions increases, problems emerge with the first law \eqref{eq:law:1}. Finally, in the discussion, we conjecture several wayouts, which constitute possible directions of future analysis.

We use the conventions adapted to the mostly plus $(-+++)$ metric signature. The sums of T-theory coupling constants $c_i+c_j$ and $c_i+c_j+c_k$ are shortened, respectively, as $c_{ij}$ and ${c_{ijk}}$.
\section{T-theory}
\label{sec:action}
T-theory is a scalar-tensor theory of gravitation, in which gravity is described by the usual metric tensor $g_{ab}$ and by a scalar field $T$. The field $T$ is assumed to determine a preferred foliation, with timelike unit-normal vector field
\be
\label{eq:u:1}
u_a=-N\nabla_aT\quad N=\left(-\nabla_aT\nabla^aT\right)^{-1/2}
\ee
where $N$ is the lapse of the foliation.

The action of T-theory is the following functional of $g_{ab}$ and $u^a$:
\be
\label{eq:action:1}
S=\int_\M\frac{d^Dx\sqrt{-g}}{16\pi G}\left[R-2\Lambda+L_u\right]
\ee
where
\begin{subequations}
\begin{align}
&L_u=-Z\indices{^a^b_c_d}\nabla_au^c\nabla_bu^d\\
&Z\indices{^{ab}_{cd}}=c_1g^{ab}g_{cd}+c_2\delta^a_c\delta^b_d+c_3\delta^b_c\delta^a_d-c_4u^au^bg_{cd}
\end{align}
\end{subequations}
and all the $c_i$'s are coupling constants. The fact that \eqref{eq:action:1} depends on $T$ only upon $u^a$ makes the theory invariant under arbitrary reparametrizations $T\to f(T)$. 

T-theory has a strong relation with \Ae\, theory. Indeed \Ae\, theory has the same action as \eqref{eq:action:1}, but it considers $u^a$ as a fundamental vector field, and it enforces the unit-timelike constraint on $u^a$ by means of a Lagrange multiplier. Conversely, one can also view T-theory as a modification of \Ae\, theory, in which the vector $u^a$ is assumed to be hypersurface orthogonal. Because of this relation, we refer to $u^a$ as the aether vector.

T-theory is also connected with the infrared limit of Ho{\v r}ava gravity. Indeed, as shown in \cite{jacobson:horava}, if we decompose the action \eqref{eq:action:1} along the constant $T$ hypersurfaces, we obtain
\begin{multline}
\label{eq:action:2}
S=\int dT\int_{\Sigma_T}\frac{d^{D-1}y\,N\sqrt{h}}{16\pi G}\left[\tilde{R}-2\Lambda+\alpha a^2\right.\\ \left.+\beta K^{ab}K_{ab}-\gamma K^2\right]
\end{multline}
where $\alpha=c_{14}$, $\beta=1-c_{13}$ and $\gamma=1+c_2$; $\{y\}$ are coordinates on $\Sigma_T$; $h$ is the determinant of the intrinsic metric $h_{ab}$ of $\Sigma_T$; $\tilde{R}$ is the Ricci scalar of $h_{ab}$, and we have defined the extrinsic curvature\footnote{The underleft arrow denotes projections of the indices on the hypersurface orthogonal to $u^a$.}
\be
\label{eq:k:1}
K_{ab}=\underleftarrow{\nabla_a u_b}
\ee
along with its trace $K$, and the aether acceleration $a_a=u^b\nabla_b u_a$.

The action \eqref{eq:action:2} contains only second order spatial derivatives. It coincides with the infrared action of Ho{\v r}ava gravity, in which operators of the full HL action with higher order spatial derivatives are suppressed by a Lorentz violating scale $\Lambda_\text{LV}$, and they are neglected. Therefore, we can also view T-theory as the covariantization of infrared HL gravity. $T$ plays the role of the preferred time and we refer to it as the khronon field. Correspondingly, T-theory is often referred in the literature as khronometric theory.

In order to derive the equations of motion (EOM) for T-theory, we must vary the action \eqref{eq:action:1} with respect to $g_{ab}$ and $T$,
\be
\label{eq:vars:1}
\delta S=\int_\M \frac{d^D x\sqrt{-g}}{16\pi G}\left[\mathbb{E}_{ab}\delta g^{ab}+2\mathbb{E}_T\delta T\right]+
\begin{pmatrix}
\text{boundary}\\
\text{terms}
\end{pmatrix}
\ee
from which we read the EOM
\begin{subequations}
\begin{align}
&\mathbb{E}_{ab}=\frac{\delta S}{\delta g^{ab}}=R_{ab}-\frac{1}{2}Rg_{ab}+\Lambda g_{ab}-T\indices{^u_{ab}}=0 \label{eq:eom:1}\\
&\mathbb{E}_T=\frac{1}{2}\frac{\delta S}{\delta T}=\nabla_a\left(N\underleftarrow{\Ae}^a\right)=0 \label{eq:eom:2}
\end{align}
\end{subequations}
where the effective stress-energy tensor of the aether is
\be
\begin{split}
T\indices{^u_{ab}}=&c_1\left(\nabla_au_m\nabla_bu^m-\nabla_mu_a\nabla^mu_b\right)+c_4a_aa_b+\\
&+\nabla_mX\indices{^m_{ab}}+\frac{1}{2}L_u\,g_{ab}+(\Ae\cdot u)u_au_b-2\underleftarrow{\Ae}_{(a}u_{b)}
\end{split}
\ee
and we defined
\be
\Ae_a=\frac{1}{2}\frac{\delta S}{\delta u^a}=c_4\,a^b\nabla_au_b+\nabla_b\left(Z\indices{^{bc}_{ad}}\nabla_cu^d\right).
\ee
$X\indices{^m_{ab}}$ is a function of $u^a$ and $g_{ab}$, whose explicit expression is not needed here (see e.g. Eq.\,(11) of \cite{pacilio_smarr}).

Notice that, in \Ae-theory, the metric EOM \eqref{eq:eom:1} is left unchanged, while the aether EOM \eqref{eq:eom:2} becomes
\be
\underleftarrow{\Ae}^a=0,
\ee
from which we also see that any hypersurface orthognal solution of \Ae\, theory is also a solution of T-theory, while the converse is not necessarily true \cite{jacobson:horava}.
\section{The Hamiltonian of T-theory}
\label{sec:hamiltonian}
\subsection{Boundary conditions and boundary terms}
\label{sub:boundary}
We want to derive the Hamiltonian of T-theory, for which we need to specify a foliation. The analysis can be performed in any foliation, but we work in the preferred foliation adapted to the khronon, as in \eqref{eq:action:2}.

The actions \eqref{eq:action:1} and \eqref{eq:action:2} are not complete, because we neglected the boundary terms, which ensure that the variational problem is well defined. Therefore we have to specify boundary conditions to fix the boundary terms. We will closely follow the analysis in \cite{mattingly:lifshitz}.

We assume that there is a past timelike boundary at $T=T_-$ and a future timelike boundary at $T=T_+$. We also assume the presence of a spacelike boundary located at a suitable notion of "spatial infinity": more precisely we assume that each preferred slice $\Sigma_T$ has an outer boundary $\mathcal{B}_T$ with spacelike unit normal $s_a$, in such a way that the whole manifold has a spacelike outer boundary $\mathcal{B}=\mathcal{B}_T\times[T_-,T_+]$. In this section we neglect the possible presence of inner boundaries, and we postpone their discussion to Section \ref{sec:uh}.

The natural boundary conditions for the metric are the Dirichlet conditions $\delta g_{ab}=0$. On the other hand, as observed in \cite{mattingly:lifshitz}, the natural boundary conditions for the khronon are the Neumann conditions $\underleftarrow{\nabla_a\delta T}=0$: they ensure that the aether vector $u^a$ remains parallel to itself, i.e. that the preferred foliation is preserved at the boundary.

The boundary terms neglected in the variation \eqref{eq:vars:1} are
\be
\label{eq:vars:2}
\begin{split}
\begin{pmatrix}
\text{boundary}\\
\text{terms}
\end{pmatrix}
&=\frac{1}{16 \pi G}\int_{\partial\M}\left[g_{ab}\nabla^m\delta g^{ab}-\nabla_a\delta g^{ma}\right.\\
&\left.+A_{ab}\delta g^{ab}+B^{ma}\underleftarrow{\nabla_a\delta T}-2N\underleftarrow{\Ae}^m\delta T\right]\epsilon_m
\end{split}
\ee
where $\partial\M=\Sigma_{T_-}\cup\Sigma_{T_+}\cup\mathcal{B}$, while $A_{ab}$ and $B^{ma}$ are tensors locally constructed out of $g_{ab}$ and $u^a$, that we do not need to specify for our purposes (see e.g. Eq. (45) of \cite{pacilio_smarr}).

Because of the aforementioned boundary conditions the third and fourth term in \eqref{eq:vars:2} vanish. The first and second terms are the same as in general relativity, and therefore they induce the Brown-York boundary term in the action \cite{brown_york}.

The last term vanishes on $\Sigma_{T_-}$ and $\Sigma_{T_+}$ because $\underleftarrow{\Ae}^m$ is parallel to the preferred slices; moreover it is also expected to vanish on $\mathcal{B}$, in accordance with the reparametrization invariance $T\to f(T)$. This is valid trivially in all the explicit solutions considered below, as they are also solutions of \Ae\, theory, i.e. they satisfy $\underleftarrow{\Ae}^a=0$ globally.

Therefore the only boundary contribution to the action comes from the Brown-York term
\be
\label{eq:action:3}
S=\int_\M \frac{d^Dx\sqrt{-g}}{16\pi G}\left[R-2\Lambda+L_u\right]+\int_{\partial\M}\eta \frac{K}{8 \pi G}\epsilon_{D-1}
\ee
where: $\eta$ is equal to +1 or -1 on the portions of $\partial\M$ that are, respectively, spacelike or timelike; $\epsilon_{D-1}$ is the induced volume element on $\partial\M$; $K$ is the trace of the extrinsic curvature of $\partial\M$, i.e. $K=\nabla_au^a$ on $T_-$ and $T_+$, while $K=\nabla_as^a$ on $\mathcal{B}$.

The full decomposition of the action \eqref{eq:action:3} on the preferred slices of constant $T$ is therefore
\begin{multline}
\label{eq:action:4}
S=\int_{T_-}^{T_+} dT\left[\int_{\Sigma_T}\frac{d^{D-1}y\,N\sqrt{h}}{16 \pi G}\left(\tilde{R}-2\Lambda+\alpha a^2\right.\right.\\ \left.\left.+\beta K^{ab}K_{ab}-\gamma K^2\right)+\oint_{\mathcal{B}_T}d^{D-2}\theta\frac{\sqrt{\sigma}\,N}{8\pi G}\tensor[^2]{K}{}\right]
\end{multline}
where $\K=\left(g^{ab}+u^au^b\right)\nabla_as_b$ is the extrinsic curvature of $\mathcal{B}_T$ viewed as an hypersurface of $\mathcal{B}$, $\{\theta\}$ are coordinates on $\B$, and $\sigma$ is the determinant of the intrinsic metric of $\mathcal{B}_T$.
\subsection{Hamiltonian decomposition}
We are ready to canonically decompose the action \eqref{eq:action:4}. First of all we define the time evolution vector as
\be
\label{eq:t:1}
T^a=\left.\left(\frac{\partial x^a}{\partial T}\right)\right|_{\vec{y}=\text{const.}}
\ee
which can be split into its "normal" and "tangential" parts as
\be
\label{eq:t:2}
T^a=Nu^a+N^a
\ee
where $N$ is the lapse introduced in \eqref{eq:u:1}, and $N^a$ is the shift.

Next, by observing that
\be
\label{eq:k:2}
K_{ab}=\frac{1}{2}\lie_uh_{ab}=\frac{\dot{h}_{ab}-2D_{(a}N_{b)}}{2N}
\ee
we define the momentum conjugate to $h_{ab}$,
\be
\label{eq:p:1}
P^{ab}=\frac{\delta L}{\delta\dot{h}_{ab}}=\frac{\sqrt{h}\left(\beta K^{ab}-\gamma Kh^{ab}\right)}{16\pi G}.
\ee
As in GR the spatial metric $h_{ab}$ is the only dynamical field, while the lapse and the shift are not dynamical. \footnote{Recall that now $T$ is a spacetime label, and therefore it does not count as a dynamical field. See \cite{jacobson:horava}.}

Then the Hamiltonian is\footnote{We use bold capital letters $\Ha$, $\Ma$, $\Ja$ for the Hamiltonian, the mass and the angular momentum, respectively.}
\be
\label{eq:h:1}
\begin{split}
\Ha&=\int_{\Sigma_T}d^{D-1}y\,P^{ab}\dot{h}_{ab}-L\\
&=\int_{\Sigma_T}\frac{d^{D-1}y}{16\pi G}\left[NH-N^aH_a\right]\\
&-\oint_{\B}\frac{d^{D-2}\theta\sqrt{\sigma}}{8\pi G}\left[N\K-N^as^bp_{ab}\right]
\end{split}
\ee
where
\begin{subequations}
\begin{align}
&H=\sqrt{h}\left(\beta K^{ab}K_{ab}-\gamma K^2-\alpha a^2-\tilde{R}+2\Lambda\right)\,\\
&H_a=-2D_bP\indices{^b_a}\,
\end{align}
\end{subequations}
and
\be
p^{ab}=\beta K^{ab}-\gamma K h^{ab}.
\ee
Eq. \eqref{eq:h:1} is the "off-shell" Hamiltonian. To obtain the "on-shell" Hamiltonian we must impose the EOM for $N$ and $N^a$. The EOM for $N^a$ is simply $H_a=0$. To compute the EOM for $N$, observe that $a_a=N^{-1}D_aN$; then, functionally deriving \eqref{eq:action:2} with respect to $N$, we get 
\be
H=-2\sqrt{h}N^{-1}D_a(\alpha D^aN)\equiv-2\sqrt{h}N^{-1}D_a(\alpha N a^a).
\ee
Therefore the on-shell Hamiltonian is
\be
\label{eq:h:2}
\Ha=-\oint_{\B}\frac{d^{D-2}\theta\sqrt{\sigma}}{8\pi G}\left[N\K+c_{14}N\as-N^as^bp_{ab}\right]
\ee
Notice that the second term in \eqref{eq:h:2} was neglected in Eq.\,(34) of \cite{mattingly:lifshitz}, while it is correctly included in Eq. (52) of \cite{jacobson_donnelly}, which deals only with asymptotically flat solutions.
\subsection{Definition of mass: dealing with divergences}
\label{sub:regularization}
The mass $\Ma$ is defined as the value of the Hamiltonian associated with asymptotic time translations. However, as described in \cite{hawking_horowitz}, a straightforward application of \eqref{eq:h:2} would result in a divergent expression, and therefore the Hamiltonian must be suitably regularized.

For example, in the case of asymptotically flat GR solutions, the so-called Hawking-Horowitz prescription reads
\be
\label{eq:h:3}
\Ma=-\oint_{\B}\frac{d^{D-2}\theta\sqrt{\sigma}}{8\pi G}N\left[\K-\K_0\right]
\ee
where the subscript $0$ means evaluation over the background solution.

Notice that this is different from
\be
\label{eq:h:4}
\Ha-\Ha_0=-\oint_{\B}\frac{d^{D-2}\theta\sqrt{\sigma}}{8\pi G}\left[N\K-N_0\K_0\right]
\ee
because subleading contributions from $N-N_0$ can combine with $\K_0$, yielding additional finite terms when integrated. 

We generalize the Hawking-Horowitz prescription in order to accommodate a nonvanishing shift at spatial infinity. This can be done as follows.

First define a local spacetime tetrad $e_a^I$, $I=0,1,2,3$, and choose the timelike member of the tetrad such that it coincides with the unit-timelike normal to the slices of the foliation: $e_a^0=u_a$.

Second, in a neighborough of $\mathcal{B}$, define the "4-current"
\be
J_a=\left[\K+c_{14}\as\right]u_a+p_{ab}s^b
\ee
in terms of which the Hamiltonian density is
\be
\label{eq:tj:1}
T^aJ_a=-N\left[\K+c_{14}\as\right]+N^as^bp_{ab}
\ee
where we have used \eqref{eq:t:2} and the fact that $u^a$ is orthogonal to $p_{ab}$.

Project both $T^a$ and $J_a$ along the tetrad interal directions
\be
\label{eq:tj:2}
T^I=T^ae^I_a\qquad J_I=J_ae^a_I.
\ee
Then we prescript to regularize the mass as
\be
\label{eq:h:5}
\Ma=\oint_{\B}\frac{d^{D-2}\theta\sqrt{\sigma}}{8\pi G}T^I\left(J_I-J_I\rvert_0\right)
\ee
where again the subscript $0$ means subtraction of the background current. For the same reason as before, this is not equivalent to $\Ha-\Ha_0$.

From now on we adopt the prescription \eqref{eq:h:5} as our definition of mass. Notice that this regularization procedure is not specific to T-theory, and it can be applied to any theory with a well-defined canonical Hamiltonian. 
\section{Universal horizons}
\label{sec:uh}
The analysis of black hole solutions in the context of \Ae\, theory and Ho{\v r}ava theory revealed the existence of a novel type of event horizons, alongside the usual Killing horizons. These new horizons, named universal horizons, act as future event horizons for modes of arbitrary speed, and therefore they are the relevant event horizons when Lorentz symmetry is broken and superluminal dispersion relations are allowed.

Universal horizons were first found in four-dimensional static asymptotically flat solutions of \Ae\, theory and T-theory \cite{blas_uh, barausse_black, mattingly:uh}. Four-dimensional static asymptotically (anti-)de Sitter universal horizons were found for a specific choice of the couplings \cite{mattingly:max}. Four-dimensional static asymptotically flat slowly rotating universal horizons in T-theory were analyzed in \cite{barausse_slowly}, while for generic choices of the couplings they were found to be absent in \Ae\, theory \cite{barausse_slowly_2}. Three-dimensional fully rotating solutions in T-theory were studied in \cite{btz_uh} in the coupling branch $c_{14}=0$. Three-dimensional static asymtptotically Lifshitz universal horizons were treated numerically in \cite{mattingly:lifshitz}. Charged static universal horizons were analyzed in \cite{ding_charged, ding_3d, ding_radiation}. Finally, static and slowly rotating universal horizons in a full (not truncated) version of Ho{\v r}ava gravity were discussed in \cite{btz_uh_uv}.

In all these cases, the UH is a leaf of the preferred foliation. A general framework to study the causal structure of spacetimes with a preferred foliation was introduced in \cite{uh_global}. It was found that, in a stationary spacetime with timelike Killing vector $\xi$, the necessary and sufficient conditions for a leaf to be a universal horizon are $\ux_\text{UH}=0$ and $\ax_\text{UH}\neq0$.

We are then led to the condition that $u_a$ be normal to the universal horizon. As discussed in \cite{mattingly:lifshitz}, this ensures that no additional boundary term is needed in the action when the inner boundary is a UH, and moreover that the contributions to the on-shell Hamiltonian \eqref{eq:h:2} from the UH vanish. Therefore the expression of the Hamiltonian is unaffected by the presence of a UH, and the definition of the mass remains the same.

The quantity $\ax$ is directly related to a peeling notion of surface gravity for universal horizons, first introduced in \cite{cropp_ray}
\be
\label{eq:ku:1}
\ku=\left.\frac{1}{2}u^a\nabla_a\ux\right|_\text{UH}\equiv\frac{1}{2}\ax_\text{UH}.
\ee
It was later proved in \cite{uh_global} that $\ax=\text{const.}\neq0$ on the UH, i.e. $\ku$ obeys a zeroth law. Moreover  it was found in \cite{mattingly_tunneling, bhatta_thesis, ding_radiation} that $\ku$ is related to a notion of tunneling temperature of the universal horizon. In particular, via a tunneling computation of pair creation at the universal horizon, the associated tunneling temperature is
\be
\label{eq:ku:2}
\tu=\left(\frac{N-1}{N}\right)\frac{\ku}{\pi},
\ee
where $N$ is the dominant UV polynomial behavior of the dispersion relation of the modes created at the UH, $\omega\sim p^N$. The modes are of course understood to be superluminal, $N>1$.

Notice however that, by means of a collapsing null shell calculation, \cite{parentani} concluded that the details of the Hawking radiation at late time are independent on the UH; moreover the late time spectrum has the characteristic Killing temperature $\kk$, at least to leading order in the small parameter $\kk/\Lambda_\text{LV}$, where $\Lambda_\text{LV}$ is the UV Lorentz violating scale entering in the modified superluminal dispersion relation. The clarification of this issue is of manifest theoretical interest. We will further comment about the role of Killing horizons in the final discussion.

Since $\tu$ depends on $N$, it induces an $N$-dependence also on $\su$ in \eqref{eq:law:1}, which in turn implies an awful species-dependence of the entropy $\su$. However, in a UV completion of Ho{\v r}ava-Lifshitz, $N$ becomes a universal constant dictated by the asymptotic Lifshitz symmetry in the UV. For definiteness we work in the limit $N\to\infty$, the case of a finite $N$ differing by just a multiplicative factor.

Now that we have clarified the definition of $\tu$, we can proceed with the study of the first law. We consider exact black hole solutions of T-theory with a universal horizon, with an increasing level of complexity. In particular, we first consider the four-dimensional static asymptotically flat UHs of \cite{mattingly:uh}; then we turn to the four-dimensional static asymptotically AdS UH obtained in \cite{mattingly:max}; finally we study the three-dimensional asymptotically AdS fully rotating UHs of \cite{btz_uh}. We do not consider the charged UHs \cite{ding_radiation, ding_3d}, whose first laws have been already analyzed in the respective papers.
\section{Study of the first law}
\label{sec:law}
\subsection{(3+1) static asymptotically flat UHs}
\label{sub:law:1}
In \cite{mattingly:uh} two exact static spherically symmetric UH solutions were derived in (3+1) spacetime dimensions. They were obtained within the two branches of the theory $c_{123}=0$ and $c_{14}=0$.

The line element has the form
\be
\label{eq:line:1}
ds^2=-e(r)dt^2+\frac{dr^2}{e(r)}+r^2d\Omega^2
\ee
while the aether vector has the form
\be
\label{eq:u:2}
u_adx^a=\ut dt-\frac{\st}{e(r)}dr
\ee
and the unit-timelike constraint $u^2=-1$ relates the three functions $e(r)$, $\ut$ and $\st$ as
\be
\label{eq:constraint:1}
\ut^2-\st^2=e(r).
\ee
Here we express the solutions in terms of the coordinates $(t,r,\theta,\phi)$. From \eqref{eq:u:2}, one can switch to the preferred frame coordinates $(T,r,\theta,\phi)$ by the transformation
\be
\label{eq:dT}
dt=dT+\frac{\st}{\ut\,e(r)}dr.
\ee
It is also convenient to introduce the vector
\be
\label{eq:s:1}
s_adx^a=\st dt-\frac{\ut}{e(r)}dr
\ee
which is unit-spacelike and orthogonal to $u^a$ everywhere. When evaluated at the spatial boundary, it coincides with $s_a$ as defined in Sec.\,\ref{sub:boundary}.

The metric \eqref{eq:line:1} possesses a timelike Killing vector
\be
\label{eq:t:3}
t^a=-\ut u^a+\st s^a\,.
\ee
The reader can verify explicitly that in the preferred frame $t^a\equiv T^a$, according to the definition \eqref{eq:t:1}. Therefore $N=-\ut$ and $N^a=\st s^a$. It then follows, from the definition \eqref{eq:tj:1} and \eqref{eq:tj:2}, and from the observation that one can choose $u^a$ and $s^a$ as two spacetime tetrads, that
\be
\label{eq:tj:3}
T^IJ_I=\ut\left[\K+c_{14}\as\right]+\st\left[\beta s^as^bK_{ab}-\gamma K\right].
\ee
To proceed further, we must specify the values of the functions $e(r)$, $\ut$ and $\st$. We adopt the parametrization given in \cite{mattingly:max},
\begin{widetext}
\be
\begin{array}{l|c|c}
\toprule
\text{} & c_{14}=0 & c_{123}=0\\
\midrule
e(r) & 1-\frac{4\ru}{3r}-\frac{c_{13}}{3(1-c_{13})}\frac{\ru^4}{r^4} & 1-\frac{2\ru}{r}-\frac{(c_{14}-2c_{13})}{2(1-c_{13})}\frac{\ru^2}{r^2}\\
\midrule
\ut & -\left(1-\frac{\ru}{r}\right)\sqrt{1+\frac{2\ru}{3r}+\frac{\ru^2}{3r^2}} & -1+\frac{\ru}{r}\\
\midrule
\st & \frac{\ru^2}{\sqrt{3(1-c_{13})}r^2} & \frac{\ru}{r}\sqrt{\frac{2-c_{14}}{2(1-c_{13})}}\\
\bottomrule
\end{array}
\ee
\end{widetext}
where in both cases $\ru$ is the radius of the universal horizon.

From a direct evaluation of \eqref{eq:tj:3} we get
\be
\label{eq:tj:4}
T^I\left(J_I-J_I\rvert_0\right)=
\begin{cases}
-\frac{4\ru}{3r^2} + O\left(r^{-3}\right) & \text{if $c_{14}=0$}\\
-2\left(1-\frac{c_{14}}{2}\right)\frac{\ru}{r^2}+O\left(r^{-3}\right) & \text{if $c_{123}=0$}
\end{cases}
\ee
and therefore
\be
\label{eq:m:1}
\Ma=\begin{cases}
\frac{2\ru}{3G} & \text{if $c_{14}=0$}\\
\left(1-\frac{c_{14}}{2}\right)\frac{\ru}{G} & \text{if $c_{123}=0$}
\end{cases}
\ee
[The leading terms in \eqref{eq:tj:4} come solely from the component of $J_I$ parallel to $u_I$. This corresponds to the fact that $u^a$ and $T^a$ are asymptotically aligned.]

Next, we evaluate $\tu$ \cite{cropp_ray}:
\be
\label{eq:tu:1}
\tu=\begin{cases}
\sqrt{\frac{2}{3(1-c_{13})}}\frac{1}{2\pi\ru} & \text{if $c_{14}=0$}\\
\sqrt{\frac{2-c_{14}}{2(1-c_{13})}}\frac{1}{2\pi\ru} & \text{if $c_{123}=0$}
\end{cases}
\ee
Combining \eqref{eq:m:1} and \eqref{eq:tu:1}, we see that a first law is satisfied in the form
\be
\label{eq:law:22}
d\Ma=\tu d\left(\alpha\frac{A_\text{UH}}{4G}\right)
\ee
where $A_\text{UH}=4\pi\ru^2$ is the area of the universal horizon, and $\alpha$ is a constant equal to
\be
\label{eq:alpha:1}
\alpha=\begin{cases}
\sqrt{\frac{2(1-c_{13})}{3}} & \text{if $c_{14}=0$}\\
\sqrt{\left(1-\frac{c_{14}}{2}\right)(1-c_{13})} & \text{if $c_{123}=0$}
\end{cases}
\ee
Therefore we are led to interpret $\alpha A_\text{UH}/4G$ as the entropy $\su$ of the universal horizon. The fact that $\alpha$ is different in the two cases is not an issue, because they are distinct branches of the theory.
\subsection{(3+1) static asymptotically AdS UHs}
\label{sub:law:2}
As it was shown in \cite{mattingly:max}, a generic feature of (anti-)\,de Sitter solutions of T-theory is that the aether vector $u^a$ and the time evolution vector $T^a$ become misaligned at infinity, unless a misalignment parameter is fine-tuned. The misalignment parameter also induces an effective cosmological constant in the metric, different from the bare one appearing in the Lagrangian: even if one starts with a negative bare cosmological constant, a positive or null effective cosmological constant is possible. In the following we the case of a negative effective cosmological constant, in such a way that the evolution vector $T^a$ is timelike at the external boundary. 

\cite{mattingly:max} further showed that (3+1)-dimensional black holes with AdS asymptotics are possible in T-theory only when $c_{14}=0$. Therefore it would be natural to fine-tune the misalignment to 0 from the very beginning, because any conclusion involving the misalignment would not have general validity beyond $c_{14}=0$. However, since our final results do not depend on the misalignment, we can easily work in the most generic case.

The solution still has the form \eqref{eq:line:1}-\eqref{eq:constraint:1}, with the functions $e(r)$, $\ut$, and $\st$ given by \cite{mattingly:max}
\begin{widetext}
\begin{subequations}
\begin{align}
&\ut=-\frac{r}{l}\left(1-\frac{\ru}{r}\right)\sqrt{1+\frac{2\ru}{r}+\frac{(3\ru^2+l^2)(\ru^2+2\ru r+3r^2)}{3r^4}}\\
&\st=\frac{r}{\lambda}+\frac{\ru^2}{r^2\sqrt{3(1-c_{13})}}\sqrt{\frac{3\ru^2+l^2}{l^2}}\\
&e(r)=1-\frac{\bar{\Lambda}r^2}{3}-\frac{r_0}{r}-\frac{c_{13}}{3(1-c_{13})}\left(\frac{3\ru^2+l^2}{l^2}\right)\frac{\ru^4}{r^4}\\
&r_0=\frac{2\ru(3\ru^2+2l^2)}{3l^2}+\frac{2\ru^2}{\lambda\sqrt{3(1-c_{13})}}\sqrt{\frac{3\ru^2+l^2}{l^2}}
\end{align}
\end{subequations}
\end{widetext}
where $\lambda$ is the misalignment parameter. The additional parameter $l$ and the effective cosmological constant $\bar{\Lambda}$ are not independent, but they depend upon $\lambda$ and the bare cosmological constant $\Lambda$ through the relations
\be
\frac{\bar{\Lambda}}{3}=\frac{\Lambda}{3}-\frac{c_{13}+3c_2}{2\lambda^2}=\frac{1}{\lambda^2}-\frac{1}{l^2}\,.
\ee
To compute the mass, observe that now $T^I\left(J_I-J_I\rvert_0\right)$ receives leading contributions also from the components of $J_I$ parallel to $s_I$. This clearly corresponds to the fact that $u^a$ and $T^a$ are misaligned at infinity. When we subtract the background $J_I\rvert_0$, we must consider $l$ and $\lambda$ as part of the maximally symmetric background, obtained by sending $\ru\to 0$. Hence the mass turns out to be
\be
\label{eq:m2}
\Ma=\frac{\ru(3\ru^2+2l^2)}{3l^2G}+\frac{\sqrt{1-c_{13}}\ru^2}{\sqrt{3}\lambda G}\sqrt{\frac{3\ru^2+l^2}{l^2}}
\ee
while $\tu$ is
\begin{multline}
\label{eq:tu:2}
\tu=\frac{1}{2\pi\ru\sqrt{1-c_{13}}}\left[\sqrt{\frac{3\ru^2+l^2}{l^2}}\right.\\ \left.+\frac{\ru\sqrt{3(1-c_{13})}}{\lambda}\right]\sqrt{\frac{9\ru^2+2l^2}{3l^2}}.
\end{multline}
Since $l$ and $\lambda$ are background quantities, to study the first law we must vary $\Ma$ only with respect to $\ru$:
\begin{multline}
\frac{\partial\Ma}{\partial\ru}=\left(\frac{9\ru^2+2l^2}{3l^2G}\right)\left[\sqrt{\frac{3\ru^2+l^2}{l^2}}\right.\\ \left.+\frac{\ru\sqrt{3(1-c_{13})}}{\lambda}\right]\left(\frac{3\ru^2+l^2}{l^2}\right)^{-1/2},
\end{multline}
from which we see that
\be
\label{eq:tdm:1}
\frac{1}{\tu}\frac{\partial\Ma}{\partial\ru}=\frac{2\pi\ru\sqrt{1-c_{13}}}{G}\sqrt{\frac{9\ru^2+2l^2}{9\ru^2+3l^2}}\,.
\ee
It is apparent that a first law in the form \eqref{eq:law:22} cannot be satisfied. If we enforce the Clausius relation
\be
\label{eq:clausius}
\frac{1}{\tu}\frac{\partial\Ma}{\partial\ru}=\frac{\partial\su}{\partial\ru}
\ee
we can solve it for $\su$, thus obtaining
\begin{multline}
\label{eq:su:1}
\su=\frac{\pi\sqrt{1-c_{13}}}{18G}\left[2\sqrt{3(2l^4+15l^2\ru^2+27\ru^4)}\right.\\
\left.-l^2\ln\left(5l^2+18\ru^2+2\sqrt{3(2l^4+15l^2\ru^2+27\ru^4)}\right)\right]
\end{multline}
modulo an integration constant, that can be chosen such that $\su=0$ when $\ru=0$. We stress that the Clausius relation \eqref{eq:tdm:1} and \eqref{eq:clausius}, and thus the result \eqref{eq:su:1}, do not depend on the value of $\lambda$.

Expression \eqref{eq:su:1} is certainly awkward. While it is known that, in the case of Killing horizons, the entropy is not always the area but it depends on the dynamics of the theory \cite{waldentropy2}, such arguments have not yet been successfully generalized to Lorentz violating theories (see \cite{foster_noether} for a first attempt and \cite{mohd_aether} for a later one).

Therefore we must be very cautious about the interpretation of expression \eqref{eq:su:1}. It might be signaling that there is something wrong with the Clausius relation \eqref{eq:clausius}, and with a naive first law in the form \eqref{eq:law:1}.
\subsection{(2+1) rotating asymptotically AdS UHs}
\label{sub:law:3}
Fully rotating BH solutions of astrophysical relevance have not yet been found in Lorentz violating theories. (3+1) asymptotically flat slowly rotating BHs were extensively studied in \cite{barausse_nogo, wang_slowly, barausse_slowly, barausse_slowly_2}; however they are not appropriate for a study of the first law, because deviations with respect to the static case occur at quadratic level in the angular momentum.
In (2+1) dimensions fully rotating BHs were found in T-theory, in the branch $c_{14}=0$ \cite{btz_uh}. They are the equivalent of the BTZ solution in GR. Universal horizons are possible in these solutions, and therefore they constitute a working arena in which to test the effects of rotation.

The line element has the form
\be
\label{eq:line:2}
ds^2=-e(r)dt^2+\frac{dr^2}{e(r)}+r^2\left(d\phi+\Omega(r)dt\right)^2
\ee
while the aether vector has still the form
\be
\label{eq:u:3}
u_adx^a=\ut dt-\frac{\st}{e(r)}dr\,.
\ee 
The unit constraint on the aether implies again the relation \eqref{eq:constraint:1}. As before it is convenient to introduce the unit-spacelike vector $s_a$ orthogonal to the aether
\be
\label{eq:s:2}
s_adx^a=\st dt-\frac{\ut}{e(r)}dr\,.
\ee
The transition to the preferred frame $(T,r,\phi)$ is again dictated by the change of variables \eqref{eq:dT}.

The line element \eqref{eq:line:2} is axisymmetric with respect to the Killing vector $\phi^a=(0,0,1)$, and possesses time translational symmetry with respect to the Killing vector
\be
\label{eq:t:4}
t^a=-\ut u^a+\st s^s+\Omega(r)\phi^a\,.
\ee
The reader can again verify that $t^a$ coincides with the preferred time evolution vector $T^a$.

The functions $e(r)$, $\Omega(r)$, $\ut$ and $\st$ are\footnote{We give them in a different parametrization with respect to the original one in \cite{btz_uh}, in which we highlight the role of $\ru$.}
\begin{subequations}
\begin{align}
&e(r)=-r_0+\frac{\bar{J}^2}{4r^2}-\bar{\Lambda}r^2 \\&\Omega(r)=-\frac{J}{2r^2}\\
&\ut=-\frac{1}{l}\left(\frac{r^2-\ru^2}{r}\right)\\&\st=\frac{r}{\lambda}+\frac{1}{r}\sqrt{\frac{\ru^4}{l^2(1-c_{13})}-\frac{J^2}{4}}
\end{align}
\end{subequations}
where
\begin{subequations}
\begin{align}
&\bar{\Lambda}=\Lambda-\frac{(2c_2+c_{13})}{\lambda^2}=\frac{1}{\lambda^2}-\frac{1}{l^2}\\
&\bar{J}^2=J^2-\frac{4c_{13}\ru^4}{l^2(1-c_{13})}\\
&r_0=\frac{2\ru^2}{l^2}+\frac{2}{\lambda}\sqrt{\frac{\ru^4}{l^2(1-c_{13})}-\frac{J^2}{4}}
\end{align}
\end{subequations}
and where $\lambda$ is the misalignment parameter.

The mass, which receives contributions from the misalignment terms, is
\be
\label{eq:m:3}
\Ma=\frac{1}{4G}\left(\frac{\ru^2}{l^2}+\frac{(1-c_{13})}{\lambda}\sqrt{\frac{\ru^4}{l^2(1-c_{13})}-\frac{J^2}{4}}\right)\,.
\ee
We also need the expression for the total angular momentum $\Ja$, which can be obtained by replacing $T^a$ with $\phi^a$ into Eqs. \eqref{eq:tj:1} and \eqref{eq:tj:2}:
\be
\label{eq:j}
\Ja=\frac{(1-c_{13})J}{8G}\,.
\ee
Finally $\tu$ is
\be
\label{eq:tu:3}
\tu=\frac{1}{l\pi\ru}\left[\frac{\ru^2}{\lambda}+\sqrt{\frac{\ru^4}{l^2(1-c_{13})}-\frac{J^2}{4}}\right]\,.
\ee
For the solution to be well defined, the constraint
\be
\label{eq:bound}
\frac{\ru^4}{l^2(1-c_{13})}-\frac{J^2}{4}\ge0
\ee
must be satisfied. When the bound \eqref{eq:bound} is saturated, an interesting fact happens in the limit $\lambda\to\infty$: the function $\st$ vanishes, and therefore $e(r)\equiv\ut^2$. In turn, this implies that the universal horizon coincides with the Killing horizon, i.e. it degenerates into a null leaf. Such a degenerate UH is not in contradiction with the discussion of Sec.\,\ref{sec:uh}: indeed from \eqref{eq:tu:3} $\ku$ vanishes as well, and therefore the condition for the UH to be a leaf does not hold anymore. The existence of degenerate UHs was first pointed out in \cite{wang:existence} for $c_{14}=0$, and in \cite{degenerate:uh} for a generic choice of the couplings. In view of these considerations, in the following we assume that  \eqref{eq:bound} holds strictly, in such a way to deal with a nondegenerate UH.

For our purposes, it is convenient to consider separately the static case $J=0$ from the rotating case $J\neq 0$.

In the static case the Clausius relation becomes
\be
\frac{\partial\su}{\partial\ru}=\frac{1}{\tu}\frac{\partial\Ma}{\partial\ru}=\frac{\pi\sqrt{1-c_{13}}}{2G}
\ee
from which it follows that\footnote{See also Eq.\,(4.26) of \cite{ding_3d}.}
\be
\label{eq:su:2}
\su=\sqrt{1-c_{13}}\frac{P_\text{UH}}{4G}
\ee
where $P_\text{UH}=2\pi\ru$ is the perimeter of the UH. As in the previous case, the Clausius relation and $\su$ do not depend on the value of $\lambda$.

The result \eqref{eq:su:2} seems promising. However the situation changes completely when we consider the rotating case: indeed it turns out the Clausius relation is not solvable at all. Let us be more explicit.

Since we are in a rotating setting, we must expect a work term in the first law \eqref{eq:law:1} of the form $\Omega_\text{UH}d\Ja$, where $\Omega_\text{UH}=-\Omega(\ru)$ is the frame dragging at the UH. From the first law
\be
\label{eq:law:2}
d\Ma=\tu d\su+\ou d\Ja
\ee
we obtain the Clausius relations
\begin{subequations}
\begin{align}
& \frac{\partial\su}{\partial\ru}=\frac{1}{\tu}\frac{\partial\Ma}{\partial\ru}\\
& \frac{\partial\su}{\partial J^2}=\frac{1}{\tu}\frac{\partial\Ma}{\partial J^2}-\frac{(1-c_{13})\ou}{16G\tu J}
\end{align}
\end{subequations}
where in the second line we took into account that $\Ja=(1-c_{13})J/8G$ and that $\ou$ is linear in $J$. By explicit computation we get
\begin{subequations}
\begin{align}
& \frac{\partial\su}{\partial\ru}=\frac{\pi\ru^2}{2G l}\left(\frac{\ru^4}{l^2(1-c_{13})}-\frac{J^2}{4}\right)^{-1/2} \label{eq:clausius:2}\\
& \frac{\partial\su}{\partial J^2}=-\frac{(1-c_{13})\pi l}{32G \ru}\left(\frac{\ru^4}{l^2(1-c_{13})}-\frac{J^2}{4}\right)^{-1/2} \label{eq:clausius:3}
\end{align}
\end{subequations}
which are again independent from $\lambda$.

Now, if we integrate \eqref{eq:clausius:3}, we obtain
\be
\label{eq:su:3}
\su=\frac{(1-c_{13})\pi l}{4G \ru} \sqrt{\frac{\ru^4}{l^2(1-c_{13})}-\frac{J^2}{4}}+f(\ru)
\ee
where $f$ is a function depending only on $\ru$. But, if we differentiate \eqref{eq:su:3} with respect to $\ru$, the result differs from \eqref{eq:clausius:2} by terms depending also on $J$, which therefore cannot be compensated by any choice of $f$. Actually the derivative of \eqref{eq:su:3} is not even proportional to \eqref{eq:clausius:2}, which shows that the problem cannot be alleviated by averaging the three terms in \eqref{eq:law:2} with three appropriate constants. Therefore we end up with a contradiction, and the Clausius relations are not integrable, as anticipated.
\section{Discussion}
\label{sec:discussion}
How do we interpret the results of Secs.\,\ref{sub:law:1}-\ref{sub:law:3}? We looked for a first law in the form \eqref{eq:law:1}. The four-dimensional asymptotically flat case is encouraging, because the proportionality between $\su$ and the area of the UH suggests that T-theory respects a form of holographic principle.

However we see that, as soon as we generalize to the AdS case, $\su$ becomes a complicated functional of $\ru$. Although this is still mathematically acceptable, we do not have any physical principle or motivation to trust such an awkward expression.

The situation becomes even worse when we turn our attention to a class of fully rotating solutions in three dimensions. In this case, while in the static configuration the holographic principle is respected, when we switch on the rotation an expression for $\su$ does not even exist.

On the top of this, we must add the similar problems highlighted in \cite{ding_radiation} in the case of charged universal horizons, in which it is suggested that a "Smarr mass", different from the total mass at infinity, must be defined to satisfy the first law. All of these evidences seem to imply that a simple version of UH mechanics, according to \eqref{eq:law:1}, is not satisfied by T-theory.

However, such a conclusion would be rather premature, for two reasons. First, $c_{14}=0$ is a corner sector of the theory, and it is not clear how our results would generalize to more generic couplings. Also, it might also be the case that higher order terms in HL would always end up introducing a nonzero $c_{14}$ via radiative corrections. In this case, setting this particular parameter to 0 in the infrared action would be inconsistent with the UV completed theory.

Second, AdS is not a natural asymptotic for HL. Indeed we expect that (a) astrophysical BHs are modeled by flat asymptotics; and (b) if we use HL as a holographic gravitational dual of a Lifshitz QFT, we should consider asymptotic Lifshitz symmetry, rather than AdS (see e.g.~\cite{lifshitz:holography,mattingly:lifshitz,Cheyne:2017bis}). 

Therefore the results of Secs.\,\ref{sub:law:2} and \ref{sub:law:3} signal problems that \emph{can} occur but, in order to see if they constitute \emph{actual} drawbacks of the theory, one must investigate what happens when more physical asymptotics are considered. For astrophysical BHs, this implies the study of fully rotating asymptotically flat (3+1)-dimensional solutions. As anticipated, such solutions have not yet been obtained.

Regarding the applications to holography, static asymptotically Lifshitz UHs in (2+1) dimensions were analyzed in \cite{mattingly:lifshitz}. It was shown that these UHs possess a first law of the form
\[ d\Ma\propto\tu dP_\text{UH}\,, \]
in analogy with their static (2+1) dimensional AdS counterparts [see Eq.\,\eqref{eq:su:2}]. Whether they are better behaved when rotation is switched on, is a matter for future research. Nonetheless, from our previous considerations, we expect the case of Lifshitz asymptotics to be indeed much more promising.

It is important, however, to consider also the possibility that problems with the first law at the UH are a general fact. In this case, a possible strategy would be to reevaluate the role of Killing horizons. This goes along the line of \cite{parentani}. For example, one possible solution to the problems of Secs.\,\ref{sub:law:2} and \ref{sub:law:3} is to give away the regularity of the UH, and to assume that it is a physical singularity. (That the UH might become a singularity in a realistic BH collapse was underlined both in \cite{blas_uh} and \cite{parentani}.) In this way the solution depends on a further free parameter,\footnote{Recall that regular \Ae\, and T-theory BH solutions are obtained by imposing the regularity of the $s_0$-horizon, i.e. the sound horizon of the spin-0 modes \cite{eling_bh, barausse_black, blas_uh}. However, when $c_{14}=0$, the $s_0$ horizon and the UH coincide.} whose freedom can then be exploited to obtain a viable first law at the Killing horizon.

Finally, it can also be that no first law exists at all, neither at the UH nor at the Killing horizon. After all, if you look at T-theory (and at Ho{\v r}ava gravity) as an effective field theory, the lacking of a first law is not a dramatic conclusion, as we do not expect fundamental laws to be respected in an approximate theory. 

Of course, for what we said above, it is clear that before embracing such nonconservative solutions, future efforts must be directed toward a more systematic analysis of fully rotating UHs, with better physically motivated asymptotics and less restricted parameter space. We hope that the present contribution stimulates further investigations along these lines in the future.
\begin{acknowledgements}
The authors are grateful to D. Mattingly, and R. Parentani, for helpful and constructive discussions.
S.L. acknowledges financial support from the John Templeton Foundation (JTF) grant \#51876.
\end{acknowledgements}
%
\end{document}